\documentclass[sigconf]{acmart}

\pdfoutput=1
\usepackage{booktabs} 
\usepackage[detect-weight=true, load-configurations=binary]{siunitx}
\DeclareSIUnit{\nothing}{\relax}
\usepackage{float}
\usepackage{enumitem}

\usepackage{balance}
\usepackage{subfig}
\usepackage{caption}
\usepackage{pgfplots}
\usepackage{pgf}
\usepackage{pgfplotstable}
\usepackage{tikz}
\usetikzlibrary{patterns}
\usetikzlibrary{snakes,arrows,shapes}
\usepackage{tikz,tkz-tab}
\usepackage{circuitikz}
\usetikzlibrary{positioning, automata, graphs, trees, fit, arrows, shapes}
\usetikzlibrary{backgrounds,patterns,matrix,calc,shadows,plotmarks, circuits.logic.US}
\usetikzlibrary{decorations}
\usepackage{textcomp}
\usepackage[noend,algoruled,linesnumbered]{algorithm2e}
\SetKwProg{Fn}{Function}{}{end}
\SetEndCharOfAlgoLine{}
\usepgfplotslibrary{groupplots}
\pgfplotsset{compat=1.9}
\usepackage{multirow}
\makeatletter
\newcommand\footnoteref[1]{\protected@xdef\@thefnmark{\ref{#1}}\@footnotemark}
\makeatother

\usepackage{makecell}

\usepackage{threeparttable}

\usepackage{etoolbox}
\makeatletter
\patchcmd\@combinedblfloats{\box\@outputbox}{\unvbox\@outputbox}{}{%
  \errmessage{\noexpand\@combinedblfloats could not be patched}%
}%
\makeatother

\usepackage{listings}
\lstdefinelanguage{p4}
{ morekeywords={*,extern_type, attribute, type, method, extern, action, control, void, parser,state, start, transition, extract, select, default, accept, out, in, inout},
  sensitive=true,
  morecomment=[l]{//}, 
  morecomment=[s]{/*}{*/}, 
  morestring=[b]" 
}

\usepackage{fixltx2e}
\usepackage[normalem]{ulem}
\useunder{\uline}{\ul}{}

\usetikzlibrary{arrows,shapes.gates.logic.US,shapes.gates.logic.IEC,calc}

\usepackage{pgfplots}
\usepackage{tikz,tkz-tab}
\usepackage{circuitikz}
\usetikzlibrary{positioning, automata, graphs, trees, fit, arrows, shapes}
\usetikzlibrary{backgrounds,patterns,matrix,calc,shadows,plotmarks, circuits.logic.US}
\usetikzlibrary{decorations}

\usepackage{lipsum}
\usepackage{setspace}
\usepackage{xfrac}

\setcopyright{acmcopyright}

\acmDOI{10.475/123_4}

\acmISBN{123-4567-24-567/08/06}

\acmConference[FPGA'18]{ACM/SIGDA International Symposium on Field-Programmable Gate Arrays}{February 2018}{Monterey, California USA} 
\acmYear{2018}
\copyrightyear{2018}

\acmArticle{4}
\acmPrice{15.00}

\DeclareSIUnit{\bit}{b}

\begin{document}
\title{P4-compatible High-level Synthesis of Low Latency \SI{100}{\giga\bit/\second} Streaming Packet Parsers in FPGAs}

\author{Jeferson Santiago da Silva, Fran\c{c}ois-Raymond Boyer and J.M. Pierre Langlois}

\orcid{1234-5678-9012}
\affiliation{%
  \institution{Polytechnique Montr\'{e}al, Canada}
}

\email{{jeferson.silva, francois-r.boyer, pierre.langlois}@polymtl.ca}

\renewcommand{\shortauthors}{J. S. da Silva et al.}

\renewcommand{\shorttitle}{P4-compatible HLS of Low Latency \SI{100}{\giga\bit/\second} Streaming Packet Parsers in FPGAs}

\begin{abstract}

Packet parsing is a key step in SDN-aware devices. Packet parsers in SDN networks need to be both reconfigurable and fast, to support the evolving network protocols and the increasing multi-gigabit data rates. The combination of packet processing languages with FPGAs seems to be the perfect match for these requirements. 

In this work, we develop an open-source FPGA-based configurable architecture for arbitrary packet parsing to be used in SDN networks. We generate low latency and high-speed streaming packet parsers directly from a packet processing program. Our architecture is pipelined and entirely modeled using templated C++ classes. The pipeline layout is derived from a parser graph that corresponds a P4 code after a series of graph transformation rounds. The RTL code is generated from the C++ description using Xilinx Vivado HLS and synthesized with Xilinx Vivado. Our architecture achieves \SI{100}{\giga\bit/\second} data rate in a Xilinx Virtex-7 FPGA while reducing the latency by 45\% and the LUT usage by 40\% compared to the state-of-the-art.
\end{abstract}

%
%

\ccsdesc[500]{Hardware~Reconfigurable logic applications}
\ccsdesc[300]{Networks~Programming interfaces}

\keywords{FPGA; packet parsers; HLS; programmable networks; P4}

\maketitle

\section{Introduction}\label{sec:intro}


The emergence of recent network applications have opened new doors to FPGA devices. 
Dataplane realization in Software-defined Networking (SDN) \cite{sdn:12} is an example \cite{Zhou:14} of such applications. In SDN networks, the data and control planes are decoupled, and they can evolve independently of each other. When new protocols are deployed in a centralized intelligent controller, new forwarding rules are compiled to the data plane element without any change to the underlying hardware. FPGAs, therefore, offer just the right degree of programmability expected by these networks, by offering fine grain programmability with sufficient and power-efficient performance.



A standard SDN forwarding element (FE) is normally implemented in a pipelined-fashion \cite{Bosshart:2013}. Incoming packets are parsed in order to extract header fields to be matched in the processing pipelines. Theses pipelines are organized as a sequence of match-action tables. In SDN FEs, a packet parser is expected to be programmable, and it can be reconfigured at run time whenever new protocols are deployed.

Recent packet processing programming languages, such as POF \cite{Song:13} and P4 \cite{Bosshart:14}, allow describing agnostic data plane forwarding behavior. Using such languages, a network programmer can specify a packet parser to indicate which header fields are to be extracted. He can as well define which tables are to be applied, and the correct order in which they will be applied.


The main focus of this work is to propose a high-level and configurable approach for packet parser generation from P4 programs. Our design follows a  configurable pipelined architecture described in C++. The pipeline layout and the header layout templates are generated by a script after the P4 compilation. 

The contributions of this paper are classified into two classes: architectural and microarchitectural. The summary of the architectural contributions of this work is listed as follows: 

\begin{itemize}[noitemsep,topsep=0pt]
\item an open-source framework for generation of programmable packet parsers\footnote{Available at \url{https://github.com/engjefersonsantiago/Vivado\_HLS}} described in a packet processing language;
\item a modular and configurable hardware architecture for streaming packet parsing in FPGAs; and
\item a graph transformation algorithm to improve the parser pipeline efficiency.
\end{itemize}

The contributions related to the microarchitectural improvements are as follows:
\begin{itemize}[noitemsep,topsep=0pt]
\item a data-bus aligned pipelined architecture for reducing the complexity in the header analysis; and
\item a lookup table approach for fast parallel barrel-shifter implementation.
\end{itemize}

The rest of this paper is organized as follows. Section~\ref{sec:related_works} presents a review of the literature, Section~\ref{sec:method} draws the methodology adopted in this work, Section~\ref{sec:results} shows the experimental results, and Section~\ref{sec:conclusion} draws the conclusions.

\section{Related Work}\label{sec:related_works}

\textbf{Packet processing languages}. The SDN \cite{sdn:12} paradigm has brought programmability to the network environment. OpenFlow \cite{of:14} is the standard protocol to implement the SDN networks. However, the OpenFlow realization \cite{Gude:08} is protocol-dependent, which limits the genericity expected in SDN.

Song \cite{Song:13} presents the POF language. POF is a protocol-agnostic packet processing language, where the user can define the behavior of the network applications. A POF program is composed of a programmable parser and match-action tables.

P4 \cite{Bosshart:14} is an emergent protocol-independent packet processing language. P4 provides a simple network dialect to describe the packet processing. The main components of a P4 program are the header declarations, packet parser state machine, match-action tables, actions, and the control program. Recently, P4 has gained adoption in both academia and industry, and this is why we have chosen P4 as the packet processing language in this work.

\textbf{Packet parsers design.} Gibb \textit{et al.} present in \cite{Gibb:2013} a methodology to design fixed and programmable high-speed packet parsers. 
However, this work did not show results for FPGA implementation.

Attig and Brebner \cite{Attig:2011} propose a \SI{400}{\giga\bit/\second} programmable parser targeting a Xilinx Virtex-7 FPGA. Their methodology includes a domain specific language to describe packet parsers, a modular and pipelined hardware architecture, and a parser compiler. The deep pipeline of this architecture allows very high throughput at expense of longer latencies. 

Ben\'{a}cek \textit{et al.} \cite{Benacek:2016} present an automatic high-speed P4-to-VHDL packet parser generator targeting FPGA devices. The packet parser hardware architecture is composed of a set of configurable parser engines \cite{Pus:2012} in a pipelined-fashion. The generated parsers achieve \SI{100}{\giga\bit/\second} for a fairly complex set of headers, however the results showed roughly 100\% overhead in terms of latency and resources consumption when compared to a hand-written VHDL implementation.

Recently, Xilinx has released the P4-SDNet translator \cite{Xilinxp42sdnet}, partially compatible with the P4\textsubscript{16} specification, that maps a P4 description to custom Xilinx FPGA logic. One particular limitation of P4-SDNet is the lack of support for variable-sized headers.

In this work, we deal with some of the pitfalls of previous works \cite{Attig:2011, Benacek:2016}, trading-off design effort, latency, performance, and resources usage. Our pipeline layout, leads to lower latencies compared to the literature \cite{Attig:2011, Benacek:2016}. Moreover, the FPGA resource consumption in terms of lookup tables (LUTs) is reduced compared to \cite{Benacek:2016}, since instead of generating each parser code we parametrize generic hand-written templated C++ classes targeted to FPGA implementation. 

\section{Design Methodology}\label{sec:method}

This section presents the methodology followed in this work. Section~\ref{sec:arch} draws the high-level architectural view. Section~\ref{sec:microarch} deals with details on microarchitectural aspects. Section~\ref{sec:pipe_gen} presents our method to generate the parser pipeline. 

\subsection{High-Level Architecture}\label{sec:arch}

A packet parser can be seen at a high-level as a directed acyclic graph (DAG), where nodes represent protocols and edges are protocol transitions. A packet parser is implemented as an abstract state machine (ASM), performing state transition evaluations at each parser state. States belonging to the path connecting the first state to the last state in the ASM compose the set of supported protocols of an FE.

\begin{figure}[t]
\center
\subfloat[][High-level packet parser pipeline layout]
{
\tikzset{%
  input/.style    = {coordinate}, 
  output/.style   = {coordinate} 
}
\tikzstyle{header} = [draw, rectangle, 
    minimum height=1.5em, minimum width=2.5em]
\tikzstyle{pipe} = [draw, rectangle, 
    minimum height= 0.75em, minimum width=5em]
\tikzstyle{mux} = [ trapezium,   draw,   
                    shape border rotate = 270,
  inner xsep=8pt, inner ysep=8pt, outer sep=0pt,
  minimum height=.1cm]
\tikzstyle{input} = [coordinate]
\tikzstyle{output} = [coordinate]

\begin{tikzpicture}[auto,>=latex',distance=1cm]
    \node [header, name=headera, anchor=center] at (0cm, 0cm){ Header A};
    \node [pipe, right of=headera, node distance=1.5cm, rotate=90] (pipea) { Register};
    \node [header, right = .5cm of pipea.south east, anchor = north west, align=center] (headerb) { Header B};
    \node [header, right = .5cm of pipea.south west,anchor=south west,  align=center] (headerc) { Header C};
    \node [mux, anchor=center] (mux) at ($(headerb.south east |- pipea.south) + (0.75cm ,0cm)$) {}; 
    \node [pipe, right of=mux, node distance=1cm, rotate=90] (pipeb) { Register};
    \node [output, right of=pipeb, node distance=.75cm] (output) {};

    \draw [draw,->, very thick] ($(headera.west)+(-0.5,0)$) -- node[below]{\rotatebox{90}{\small Data In}} (headera.west);
    \draw [->, very thick] (headera) -- node {} (pipea);
    \draw [->, very thick] (pipea) --+(0.5cm,0) |- node {} (headerb.west);
    \draw [->, very thick] (pipea) --+(0.5cm,0) |- node {} (headerc.west);
    \fill ($(pipea)+(0.5cm,0)$) circle (2pt);
    \draw [->, very thick] (headerb.east) --+(0.15cm,0)|- node {} (mux.north west);
    \draw [->, very thick] (headerc.east) --+(0.15cm,0)|- node {} (mux.south west);
    \draw [->, very thick] (mux) -- node  {} (pipeb);
    \draw [->, very thick] (pipeb) -- node[below]{\rotatebox{90}{\small Data Out}}(output);
\end{tikzpicture}
\label{fig:high_level:pipe}
}\hfill
\subfloat[][Internal header block architecture]
{
\tikzset{%
  input/.style    = {coordinate}, 
  output/.style   = {coordinate} 
}
\tikzstyle{box} = [draw, rectangle, 
    minimum height=1.5em, minimum width=9em]
\tikzstyle{pipe} = [draw, rectangle, 
    minimum height= 0.75em, minimum width=2.5em]
\tikzstyle{mux} = [ trapezium,   draw,   
                    shape border rotate = 270,
  inner xsep=8pt, inner ysep=8pt, outer sep=0pt,
  minimum height=.1cm]
\tikzstyle{no} = [draw, color= black, circle, 
    radius=.1cm]
\tikzstyle{input} = [coordinate]
\tikzstyle{output} = [coordinate]

\begin{tikzpicture}[auto,>=latex',distance=1cm]
    \node [box, align=center, node distance=1.25cm] (pipeline) at (0cm, 0cm) { Pipeline Alignment}; 
    \node [input, name=input, left of=pipeline, node distance=2.5cm] {}; 
    \node [box, above of=pipeline, align=center, node distance=1cm] (state) { State  Transition};
    \node [input, anchor=center] (state_input) at ($(input.east |- state.175)$) {}; 
    \node [box, below of=pipeline, align=center, node distance=1cm] (extraction) { Header Extraction};
    \node [output, right of=pipeline, node distance=2.5cm] (output) {};
    \node [output, below of=extraction, node distance=2.5	cm] (phv_output) {};
    \node [output, right of=state, node distance=2.5cm] (state_output) {};

    \draw [draw,->, very thick] (input) -- node[label={[xshift=-0.55cm, yshift=0.cm]\small Data In}] {} (pipeline);
    \draw [draw,->, very thick] (state_input) -- node[label={[xshift=-0.65cm, yshift=0.cm]\small NHeader In}] {} (state.175);
    \draw [<->, very thick	] (state.-175) --+(-0.35cm,0) |- node {} (extraction.west);
    \fill ($(pipeline.west)+(-0.35cm,0)$) circle (2pt);
    \draw [->, very thick	] (pipeline) -- node[label={[xshift=0.50cm, yshift=0.cm]\small Data Out}] {}(output);
    \draw [->, very thick	] (state) -- node[label={[xshift=0.75cm, yshift=0.cm]\small NHeader Out}] {}(state_output);
    \draw [->	] (state.south) -- node[label={[xshift=-1.5cm, yshift=-0.0cm]\scriptsize Header Valid}] {}(pipeline.north);
    \draw [->	] (state.south) |-+(1.55cm,-0.25cm) |-+ (1.55cm,-1.25cm) -| node[] {}(extraction.35);
    \fill ($(state.south)+(0,-0.25cm)$) circle (1pt);
    \draw [->, very thick	] (extraction.north) -- node[label={[xshift=-0.0cm, yshift=-0.0cm]\scriptsize Header Size}] {}(pipeline.south);
    \draw [draw,->, very thick] (0,2) -- node[label={[xshift=-0.0cm, yshift=0.cm]\small Header Layout}] {} (0,1.5);
    \draw [draw,->, very thick] (extraction.south) -- node[label={[xshift=-0.0cm, yshift=0.cm]\small PHV}] {} ($(extraction.south)+(0,-0.5cm)$);

    \draw [thin, dashed](-1.9, -1.5) rectangle (1.9,1.5);
\end{tikzpicture}
\label{fig:high_level:header}
}
\caption{High-level architecture}
\label{fig:high_level}
\end{figure}
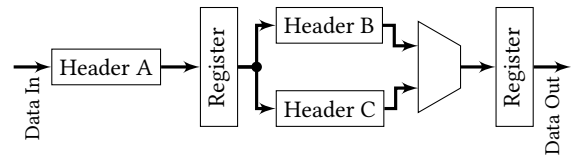
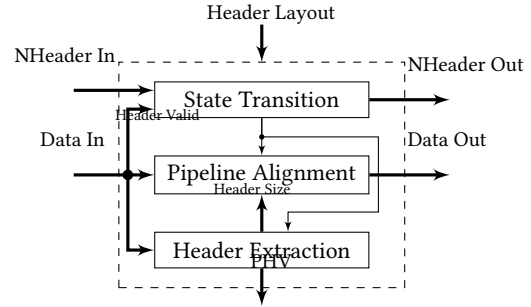

Figure~\ref{fig:high_level:pipe} depicts the high-level view of the packet parser realization proposed in this work. The proposed architecture is a streaming packet parser, requiring no packet storage. Header instances are organized in a pipelined-fashion. Headers that share the same previous states are processed in parallel. Throughout this work, we say that those headers belong to the same parser (graph) level. The depth of the parser pipeline is the length of the longest path in the parser graph. For sake of standardization, thick arrows in the figures throughout this work indicate buses, while thin arrows represent single signals.

The internal header block architecture is shown in Figure~\ref{fig:high_level:header}. This block was carefully described using templated C++ classes to offer the right degree of configurability required by the most varied set of protocol headers this architecture is intended to support. This design choice was also taken to improve bit-accuracy by accordingly setting arbitrary integer variables, reducing FPGA resources usage.

In Figure~\ref{fig:high_level:header}, the \textit{Header Layout} is a configuration parameter. It is a set of data structures required to initialize the processing objects. It includes key match offsets and sizes for protocol matching, lookup tables to determine data shift values, expressions to determine the header size, last header indication, and so forth. \textit{Data In} is a data structure that contains the incoming data to be processed in a header instance. It is composed of the data bus to be analyzed and some metadata. These metadata include data start and finish information for a given packet and packet identifier. The packet identifier is used to keep track of the packet throughout the processing pipeline and to identify which headers belong to the same packet. \textit{NHeader In} is assigned by the previous header instance indicating which is the next header to be processed. \textit{PHV} is a data structure containing the extracted fields. It includes the extracted data, number of bits extracted, a data valid information, and header and packet identifier. Signals labelled with \textit{In} and \textit{Out} are mirrored, which means that \textit{In} signals undergo modifications before being forwarded to \textit{Out}.

Internal sub-blocks execute in parallel with minimum data dependency. In fact, only the \textit{Header Valid} information must propagate among the blocks within the same clock cycle and it is generated from a basic combinational logic. \textit{Header Size} also transits from the \textit{Header Extraction} to the \textit{Pipeline Alignment} module. However, this information is only required in the next cycle, which does not constitute a true data hazard.


\subsection{Microarchitectural Aspects}\label{sec:microarch}

This subsection presents microarchitectural aspects of our proposed method. We start by presenting the state transition block. Details of the header extraction module are drawn followed by the pipeline alignment block. Then, we present the case of variable-sized headers.

\makeatletter
\pgfdeclareshape{dff}{
  \savedanchor\northeast{%
    \pgfmathsetlength\pgf@x{\pgfshapeminwidth}%
    \pgfmathsetlength\pgf@y{\pgfshapeminheight}%
    \pgf@x=0.5\pgf@x
    \pgf@y=0.5\pgf@y
  }
  \savedanchor\southwest{%
    \pgfmathsetlength\pgf@x{\pgfshapeminwidth}%
    \pgfmathsetlength\pgf@y{\pgfshapeminheight}%
    \pgf@x=-0.5\pgf@x
    \pgf@y=-0.5\pgf@y
  }
  \inheritanchorborder[from=rectangle]

  \anchor{center}{\pgfpointorigin}
  \anchor{north}{\northeast \pgf@x=0pt}
  \anchor{east}{\northeast \pgf@y=0pt}
  \anchor{south}{\southwest \pgf@x=0pt}
  \anchor{west}{\southwest \pgf@y=0pt}
  \anchor{north east}{\northeast}
  \anchor{north west}{\northeast \pgf@x=-\pgf@x}
  \anchor{south west}{\southwest}
  \anchor{south east}{\southwest \pgf@x=-\pgf@x}
  \anchor{text}{
    \pgfpointorigin
    \advance\pgf@x by -.5\wd\pgfnodeparttextbox%
    \advance\pgf@y by -.5\ht\pgfnodeparttextbox%
    \advance\pgf@y by +.5\dp\pgfnodeparttextbox%
  }

  \anchor{D}{
    \pgf@process{\northeast}%
    \pgf@x=-1\pgf@x%
    \pgf@y=.5\pgf@y%
  }
  \anchor{CLK}{
    \pgf@process{\northeast}%
    \pgf@x=-1\pgf@x%
    \pgf@y=-.66666\pgf@y%
  }
  \anchor{CE}{
    \pgf@process{\northeast}%
    \pgf@x=-1\pgf@x%
    \pgf@y=-0.33333\pgf@y%
  }
  \anchor{Q}{
    \pgf@process{\northeast}%
    \pgf@y=.5\pgf@y%
  }
  \anchor{Qn}{
    \pgf@process{\northeast}%
    \pgf@y=-.5\pgf@y%
  }
  \anchor{R}{
    \pgf@process{\northeast}%
    \pgf@x=0pt%
  }
  \anchor{S}{
    \pgf@process{\northeast}%
    \pgf@x=0pt%
    \pgf@y=-\pgf@y%
  }
  \backgroundpath{
    \pgfpathrectanglecorners{\southwest}{\northeast}
    \pgf@anchor@dff@CLK
    \pgf@xa=\pgf@x \pgf@ya=\pgf@y
    \pgf@xb=\pgf@x \pgf@yb=\pgf@y
    \pgf@xc=\pgf@x \pgf@yc=\pgf@y
    \pgfmathsetlength\pgf@x{1ex} 
    \advance\pgf@ya by \pgf@x
    \advance\pgf@xb by \pgf@x
    \advance\pgf@yc by -\pgf@x
    \pgfpathmoveto{\pgfpoint{\pgf@xa}{\pgf@ya}}
    \pgfpathlineto{\pgfpoint{\pgf@xb}{\pgf@yb}}
    \pgfpathlineto{\pgfpoint{\pgf@xc}{\pgf@yc}}
    \pgfclosepath

    \begingroup
    \tikzset{flip flop/port labels} 
    \tikz@textfont

    \pgf@anchor@dff@D
    \pgftext[left,base,at={\pgfpoint{\pgf@x}{\pgf@y}},x=\pgfshapeinnerxsep]{\raisebox{-0.75ex}{D}}

    \pgf@anchor@dff@CE

    \pgf@anchor@dff@Q
    \pgftext[right,base,at={\pgfpoint{\pgf@x}{\pgf@y}},x=-\pgfshapeinnerxsep]{\raisebox{-.75ex}{Q}}

    \pgf@anchor@dff@Qn

    \pgf@anchor@dff@R

    \pgf@anchor@dff@S
     \pgftext[bottom,at={\pgfpoint{\pgf@x}{\pgf@y}},y=\pgfshapeinnerysep]{E}
    \endgroup
  }
}

%
%

\tikzset{flip flop/port labels/.style={font=\ttfamily\scriptsize}}
\tikzset{add font/.code={\expandafter\def\expandafter\tikz@textfont\expandafter{\tikz@textfont#1}}} 
\tikzset{every dff node/.style={draw,minimum width=.6666cm,minimum 
		height=1.0cm,inner sep=1mm,outer sep=0pt,cap=round,add 
		font=\ttfamily}}
\tikzstyle{dataBus} = [draw, very very thick]
\tikzstyle{fils} = [draw]

\makeatother

\tikzstyle{decision} = [diamond, draw, fill=blue!20, 
text width=4.5em, text centered, node distance=3cm, inner sep=0pt]
\tikzstyle{block} = [rectangle, draw, fill=blue!20, 
text width=7em, text centered, rounded corners, minimum height=3em]
\tikzstyle{line} = [draw, -latex']
\tikzstyle{cloud} = [draw, ellipse,fill=red!20, node distance=3cm,
minimum height=2em]

\subsubsection{State Transition Block}\label{sec:state_tran}

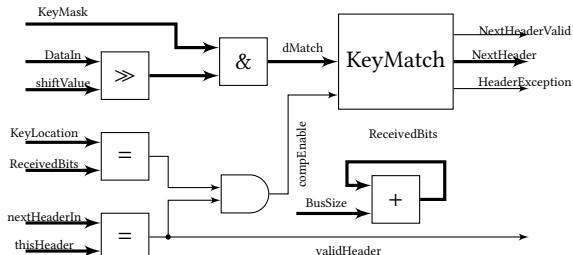
\begin{figure}[!t]
\center
\tikzset{%
  input/.style    = {coordinate}, 
  output/.style   = {coordinate} 
}
\tikzstyle{box} = [draw, rectangle, 
    minimum height=2em, minimum width=2em]
\tikzstyle{no} = [draw, color= black, circle, 
    radius=.1cm]
\tikzstyle{input} = [coordinate]
\tikzstyle{output} = [coordinate]

\begin{tikzpicture}[auto,>=latex',distance=1cm]
    \node [box, align=center, node distance=1.0cm] (shift) {$\gg$};
    \node [input, name=input, left = of shift.150, node distance=.5cm] {};
    \node [input, name=shiftVal, left = of shift.-150, node distance=.5cm] {};

    \node [box, node distance=1.0cm, draw, anchor=-150] (and1) at ($(shift.center |- shift.center) + (1.25cm, 0.cm)$) {$\&$} ;
    \node [input, name=mask, node distance=0.5cm] at ($(input.east |- input.center) + (0cm,.5cm)$) {};

    \node [box, align=center, node distance=1.0cm] (valid) at ($(shift.center |- shift.south) + (0cm,-.75cm)$) {$=$};
    \node [input, name=recB, left = of valid.-150, node distance=.5cm] {};
    \node [input, name=KeyLoc, left = of valid.150, node distance=.5cm] {};

    \node [box, align=center, node distance=1.0cm] (hvalid) at ($(valid.center |- valid.south) + (0cm,-0.75cm)$) {$=$};
    \node [input, name=thisHeader, left = of hvalid.-150, node distance=.5cm] {};
    \node [input, name=nextHeaderIn, left = of hvalid.150, node distance=.5cm] {};

    \node [box, align=center, node distance=1.0cm] (counter) at ($(hvalid.center |- hvalid.center) + (3.6cm,.5cm)$) {$+$};
    \node [input, name=recBC, left = of counter.150, node distance=.5cm] {};
    \node [input, name=BusS, left = of counter.-150, node distance=.5cm] {};

    \node [and gate US, node distance=1.0cm, draw, logic gate inputs=nn] (and2) at ($(and1.center |- hvalid.north) + (0cm,.25cm)$) {} ;
    
    \node [box, align=center, node distance=1.0cm, minimum height=4em, minimum width=4em] (cmp) at ($(counter.center |- and1.center) + (0cm,0cm)$) {KeyMatch};
    \node [output, name=phv, right = of cmp, node distance=.5cm] {};
    \node [output, name=NextHeaderValid, node distance=.5cm] at ($(phv.east |- cmp.25) + (0cm,0cm)$){};
    \node [output, name=HeaderException, node distance=.5cm] at ($(phv.east |- cmp.-25) + (0cm,0cm)$){};

    \node [output, right of=shift, node distance=0.5cm] (output) {};
    \node [output, right of=shift, node distance=0.cm] (vheader) at ($(phv.center |- hvalid.center) + (0cm,0cm)$) {};

    \draw [draw,->, very thick] (input) -- node[label={[xshift=-0.cm, yshift=-0.35cm]\tiny DataIn}] {} (shift.150);
    \draw [draw,->, very thick] (shiftVal) -- node[label={[xshift=-0.cm, yshift=-0.35cm]\tiny shiftValue}] {} (shift.-150);

    \draw [draw,->, very thick] (shift.east) --+ (.25cm,-0cm)|- node[] {} (and1.-150);
    \draw [draw,->, very thick] (mask)  --+ (2.0cm,-0cm)|-  node[label={[xshift=-1.5cm, yshift=0cm]\tiny KeyMask}]{} (and1.150);
    \draw [draw,->, very thick] (recB) --  node[label={[xshift=-0.25cm, yshift=-0.35cm]\tiny ReceivedBits}]{} (valid.-150);
    \draw [draw,->, very thick] (KeyLoc) --  node[label={[xshift=-0.25cm, yshift=-0.35cm]\tiny KeyLocation}]{} (valid.150);

    \draw [draw,->, very thick] (counter) -|+ (0.65cm,0.5cm) -|+ (-0.65cm,0.5cm) |- node[label={[xshift=0.75cm, yshift=0.25cm]\tiny ReceivedBits}]{} (counter.150);
    \draw [draw,->, very thick] (BusS) --  node[label={[xshift=-0.1cm, yshift=-0.3cm]\tiny BusSize}]{} (counter.-150);

    \draw [draw,->, very thick] (thisHeader) --  node[label={[xshift=-0.25cm, yshift=-0.35cm]\tiny thisHeader}]{} (hvalid.-150);
    \draw [draw,->, very thick] (nextHeaderIn) --  node[label={[xshift=-0.25cm, yshift=-0.35cm]\tiny nextHeaderIn}]{} (hvalid.150);
    \draw [draw,->] (valid.east) --+ (.25cm,-0cm)|- node[] {} (and2.input 1);
    \draw [draw,->] (hvalid.east) --+ (.25cm,-0cm)|- node[] {} (and2.input 2);
    \fill ($(hvalid.east)+(.25cm,-0cm)$) circle (1pt);

    \draw [draw,->, very thick] (and1) -- node[] {\tiny dMatch} (cmp);
    \draw [draw,->] (and2.output) --+ (.25cm,-0cm) |- node[pos=0.2, sloped, below] {\tiny compEnable} (cmp.-150);
    
    \draw [draw,->, very thick] (cmp) -- node[label={[xshift=0.175cm, yshift=-0.35cm]\tiny NextHeader}] {} (phv);
    \draw [draw,->] (cmp.25) -- node[label={[xshift=0.45cm, yshift=-0.35cm]\tiny NextHeaderValid}] {} (NextHeaderValid);
    \draw [draw,->] (cmp.-25) -- node[label={[xshift=0.45cm, yshift=-0.35cm]\tiny HeaderException}] {} (HeaderException);

    \draw [draw,->] ($(hvalid.east)+(.25cm,-0cm)$) -- node[below] {\tiny validHeader} (vheader); 

\end{tikzpicture}
\caption{Station transition block}
\label{fig:state}
\end{figure}

Figure~\ref{fig:state} shows the state transition block which implements part of the ASM that represents the whole parser. Each \textit{state} (header) of this ASM performs state transition evaluations by observing a specific field in the header and matching against a table storing the supported next headers for a given state. In this work, this table is filled at compilation time and it is part of what we call \textit{Header Layout}.

The state transition block uses only barrel-shifters, counters, and comparators to perform state evaluations. Such operations can be easily done in an FPGA within a single clock cycle.

In Figure~\ref{fig:state}, \textit{validHeader} is the result of a comparison between the \textit{nextHeaderIn} and \textit{thisHeader}. \textit{thisHeader} is hardwired and it is part of the header layout. \textit{validHeader} is used as an enable signal for all stateful components in the header instance. \textit{ReceivedBits} is a counter that keeps track of the number of bits received in the same header. This information is used to check if the current data window belongs to the same window in which the \textit{KeyValue} is placed in (\textit{KeyLocation}). A barrel-shifter is used to shift the input data and to align it with the \textit{KeyValue}. The bitwise AND ($\&$) operation after the barrel-shifter guarantees this alignment. Finally, the \textit{KeyMatch} compares the key aligned input data and the key table. If a match is found, the \textit{NextHeader} is assigned to the value corresponding to the match and the \textit{NextHeaderValid} is set. \textit{HeaderException} is asserted otherwise.


\subsubsection{Header Extraction Block}\label{sec:extract}

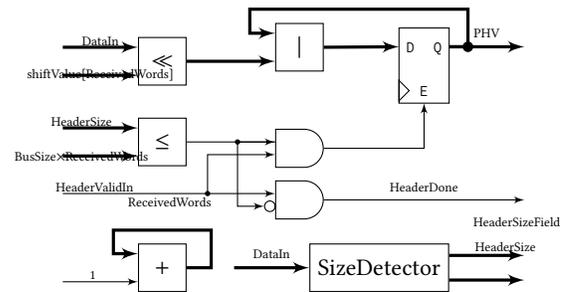
\begin{figure}[t]
\center
\tikzset{%
  input/.style    = {coordinate}, 
  output/.style   = {coordinate} 
}
\tikzstyle{box} = [draw, rectangle, 
    minimum height=2em, minimum width=2em]
\tikzstyle{no} = [draw, color= black, circle, 
    radius=.1cm]
\tikzstyle{input} = [coordinate]
\tikzstyle{output} = [coordinate]

\begin{tikzpicture}[auto,>=latex',distance=1cm]
    \node [box, align=center, node distance=1.0cm] (shift) {$\ll$};
    \node [input, name=input, left = of shift.150, node distance=.5cm] {};
    \node [input, name=shiftVal, left = of shift.-150, node distance=.5cm] {};

    \node [box, node distance=1.0cm, draw, anchor=-150] (or1) at ($(shift.center |- shift.center) + (1.5cm, 0.cm)$) {$|$} ;
	\node [dff, anchor=d, right = of or1.-37.5] (Reg) {};
    
    \node [box, align=center, node distance=1.0cm] (valid) at ($(shift.center |- shift.south) + (0cm,-.75cm)$) {$\leq$};
    \node [input, name=recB, left = of valid.-150, node distance=.5cm] {};
    \node [input, name=KeyLoc, left = of valid.150, node distance=.5cm] {};
    \node [input, name=mask, node distance=0.5cm] at ($(recB.west |- recB.center) + (0cm,-0.5cm)$) {}; 

    \node [box, align=center, node distance=1.0cm] (counter) at ($(valid.center |- valid.south) + (0cm,-1.35cm)$) {$+$};
    \node [input, name=recBC, left = of counter.150, node distance=.5cm] {};
    \node [input, name=BusS, left = of counter.-150, node distance=.5cm] {};

    \node [box, align=center, node distance=1.0cm, minimum height=2em, minimum width=2em, anchor=east] (size) at ($(Reg.east |- counter.center) + (0cm,0cm)$) {SizeDetector};
    \node [input, name=din, left = of size, node distance=.5cm] {};
    \node [input, name=hsizefield, right = of size.10, node distance=.5cm] {};
    \node [input, name=hsize, right = of size.-10, node distance=.5cm] {};

    \node [and gate US, node distance=1.0cm, draw, logic gate inputs=nn, anchor=input 1] (and2) at ($(or1.west |- valid.center) + (0cm,.0cm)$) {} ;

    \node [and gate US, node distance=1.0cm, draw, logic gate inputs=ni, anchor=input 1] (and3) at ($(or1.west |- mask.center) + (-0.cm,0cm)$) {} ;
   
    \node [output, name=phv, right = of Reg.Q, node distance=.5cm] {};
    \node [output, node distance=.5cm] (hdone) at ($(phv.east |- and3.center) + (-0.cm,0cm)$) {} ;

    \node [output, right of=shift, node distance=0.5cm] (output) {};
    \node [output, right of=shift, node distance=0.cm] (vheader) at ($(phv.center |- hvalid.center) + (0cm,0cm)$) {};

    \draw [draw,->, very thick] (input) -- node[label={[xshift=-0.cm, yshift=-0.35cm]\tiny DataIn}] {} (shift.150);
    \draw [draw,->, very thick] (shiftVal) -- node[label={[xshift=-0.cm, yshift=-0.45cm]\tiny shiftValue[ReceivedWords]}] {} (shift.-150);

    \draw [draw,->, very thick] (shift.east) --+ (.25cm,-0cm)|- node[] {} (or1.-150);
    
    \draw [draw,->, very thick] (or1.east) -- node[] {} (Reg.D);
    \draw [draw,->, very thick] (Reg.Q) -|+ (0.25cm,0.5cm) -|+ (-2.65cm,0.5cm) |- node[label={[xshift=0.75cm, yshift=0.25cm]}]{} (or1.150);    
    \fill ($(Reg.Q)+(0.25cm,0cm)$) circle (2pt);
    
    \draw [draw,->, very thick] (recB) --  node[label={[xshift=-0.25cm, yshift=-0.45cm]\tiny BusSize$\times$ReceivedWords}]{} (valid.-150);
    \draw [draw,->, very thick] (KeyLoc) --  node[label={[xshift=-0.25cm, yshift=-0.35cm]\tiny HeaderSize}]{} (valid.150);

    \draw [draw,->, very thick] (counter) -|+ (0.65cm,0.5cm) -|+ (-0.65cm,0.5cm) |- node[label={[xshift=0.75cm, yshift=0.25cm]\tiny ReceivedWords}]{} (counter.150);
    \draw [draw,->] (BusS) --  node[label={[xshift=-0.1cm, yshift=-0.3cm]\tiny 1}]{} (counter.-150);

    \draw [draw,->] (valid.east) --+ (.25cm,-0cm)|- node[] {} (and2.input 1);
    \fill ($(and2.input 1)+(-.5cm,-0cm)$) circle (1pt);

    \draw [draw,->] ($(mask)+(1.925cm,-0cm)$) -- node[] {} (and3.input 1); 
    \draw [draw,->] (and2.input 1) --+ (-.5cm,-0cm)|- node[] {} (and3.input 2);
    \draw [draw,->] (and2.output) -| node[] {} (Reg.S);

    \draw [draw,->] (mask) --+ (1.925cm,-0cm)|-  node[label={[xshift=-1.5cm, yshift=-0.825cm]\tiny HeaderValidIn}]{} (and2.input 2);
    \fill ($(mask)+(1.925cm,-0cm)$) circle (1pt);

    \draw [draw,->, very thick] (Reg.Q) -- node[] {\tiny PHV} (phv);
    \draw [draw,->] (and3.output) -- node[] {\tiny HeaderDone} (hdone);

    \draw [draw,->, very thick] (din) -- node[] {\tiny DataIn} (size);
    \draw [draw,->, very thick] (size.-10) --node[label={[xshift=0.25cm, yshift=-0.0cm]\tiny HeaderSize}]{} (hsize);
    \draw [draw,->, very thick] (size.10) --node[label={[xshift=0.4cm, yshift=-0.0cm]\tiny HeaderSizeField}]{} (hsizefield);
 
\end{tikzpicture}
\caption{Header extraction block}
\label{fig:header_extraction}
\end{figure}

Figure~\ref{fig:header_extraction} shows the header extraction block which retrieves the header information from a raw input data stream. Similarly to the state transition block, this module is implemented using barrel-shifters, comparators, and counters. Additionally, this module calculates header sizes derived from the raw input data in case of variable-sized headers. For fixed-sized headers, the header size information is hardwired at compile time.

In the header extraction module architecture, the counter \textit{ReceivedWords} is used to delimit the header boundaries for comparison with the \textit{HeaderSize}. It is also used to index a table that stores the shift amounts for the barrel-shifter. This table is fixed and it is filled at compile time. The bitwise OR ($|$) acts as an accumulator, receiving the current shifted and value accumulating it with the results from previous cycles. \textit{HeaderDone} indicates that a header has been completely extracted.

The \textit{SizeDetector} sub-block is hardwired for fixed-sized headers. For variable-sized headers, this sub-block has a behavior similar to the state transition module, returning the header size and the value of the field corresponding to the header size. More details regarding variable-sized headers are drawn in Section~\ref{sec:var_size}.


\subsubsection{Pipeline Alignment Block}\label{sec:pipe_adj}

\begin{figure}[ht]
\center
\tikzset{%
  input/.style    = {coordinate}, 
  output/.style   = {coordinate} 
}
\tikzstyle{box} = [draw, rectangle, 
    minimum height=2em, minimum width=2em]
\tikzstyle{no} = [draw, color= black, circle, 
    radius=.1cm]
\tikzstyle{input} = [coordinate]
\tikzstyle{output} = [coordinate]
\tikzstyle{mux} = [ trapezium,   draw,   
                    shape border rotate = 270,
  inner xsep=8pt, inner ysep=8pt, outer sep=0pt,
  minimum height=.1cm]

\begin{tikzpicture}[auto,>=latex',distance=1cm]
	\node [dff, anchor=center, node distance=1.0cm] (Reg) {};
    \node [box, align=center, anchor=center, node distance=1.0cm] (shift) at ($(Reg.center |- Reg.Q) + (1.5cm, 0.2cm)$) {$\ll$} ;

    \node [input, name=input, left = of Reg.D, node distance=.5cm] {};
    \node [input, name=shiftVal, left = of shift.-150, node distance=.5cm] {};

    \node [box, node distance=1.0cm, draw, anchor=center] (or1) at ($(shift.center |- shift.south) + (1.5cm, -0.35cm)$) {$|$} ;
    
    \node [box, align=center, node distance=1.0cm] (shiftr) at ($(shift.center |- shift.south) + (0cm,-1.cm)$) {$\gg$};

    \node [mux, anchor=north west] (mux) at ($(or1.center |- or1.center) + (1.25cm ,-0cm)$) {}; 
    \node [input, name=sel, below = of mux.south, node distance=.5cm] {};
    \node [box, align=center, node distance=1.0cm, minimum height=2em, minimum width=2em, anchor=center] (amount) at ($(shift.center |- Reg.center) + (0cm,-2.cm)$) {ShiftAmount};
    \node [input, name=headerSizeField, left = of amount.170, node distance=.5cm] {};
    \node [input, name=headerSize, left = of amount.-170, node distance=.5cm] {};
    \node [input, name=lShiftAmount, right = of amount.10, node distance=.5cm] {};
    \node [input, name=rShiftAmount, right = of amount.-10, node distance=.5cm] {};
    \node [input, name=lShiftAmount1, left = of shift.150, node distance=.5cm] {};
    \node [input, name=rShiftAmount1, left = of shiftr.-150, node distance=.5cm] {};
    \node [input, name=out, right = of mux.east, node distance=.5cm] {};
    \node [input, name=buss, below = of amount, node distance=.25cm] {};

    \draw [draw,->, very thick] (input) -- node[label={[xshift=-0.cm, yshift=0.cm]\tiny DataIn}] {} (Reg.D);
    \fill ($(input)+(0.5cm,-0cm)$) circle (2pt);
    \draw [draw,->, very thick] (Reg.Q) -- node[label={[xshift=-0.cm, yshift=0.35cm]\tiny}] {} (shift.-150);
    \draw [draw,->, very thick] (shift) --+ (.5cm,-0cm)|- node[] {} (or1.150);
    \draw [draw,->, very thick] (input) --+ (.5cm,-0cm)|- node[] {} (shiftr.150);
    \fill ($(input |- shiftr.150)+(0.5cm,-0cm)$) circle (2pt);
    \draw [draw,->, very thick] (shiftr) --+ (.5cm,-0cm)|- node[] {} (or1.-150);
    \draw [draw,->, very thick] (or1) -- node[] {} (mux.north west);
    \draw [draw,->, very thick] (input) -|+(.5cm,-1.75cm) --+ (4.5cm,-1.75cm) |- node[] {} (mux.south west);
    \draw [draw,->] (sel) -- node[pos=0.2, below, sloped]{\tiny HeaderValidIn} (mux.south);

    \draw [draw,->, very thick] (headerSizeField) -- node[pos=0.1, below, sloped]{\tiny headerSizeField} (amount.170);
    \draw [draw,->, very thick] (headerSize) -- node[pos=0.2, below, sloped]{\tiny headerSize} (amount.-170);
    \draw [draw,->, very thick] ($(amount.south)+(0cm,-0.5cm)$) -- node[pos=0.5, below, sloped]{\rotatebox{270}{\tiny BusSize}} (amount.south);
    \draw [draw,<-, very thick] (lShiftAmount) -- node[pos=0.1, below, sloped]{\tiny leftShiftAmount} (amount.10);
    \draw [draw,<-, very thick] (rShiftAmount) -- node[pos=0.05, below, sloped]{\tiny rightShiftAmount} (amount.-10);

    \draw [draw,->, very thick] (lShiftAmount1) -- node[pos=0.1, above, sloped]{\tiny leftShiftAmount} (shift.150);
    \draw [draw,->, very thick] (rShiftAmount1) -- node[pos=0.1,below, sloped]{\tiny rightShiftAmount} (shiftr.-150);

    \draw [draw,->, very thick] (mux) -- node[pos=0.5,above, sloped]{\tiny DataOut} (out);

%
%
%
     
\end{tikzpicture}
\caption{Pipeline alignment block}
\label{fig:pipe_align}
\end{figure}
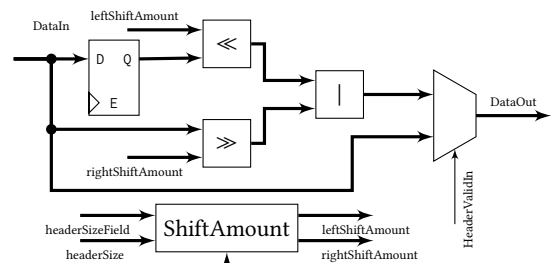

Unlike previous works, we opt for a bus-aligned pipeline architecture. That means that each stage in the parser pipeline aligns the incoming data stream before sending it to the next stage. This design choice reduces the complexity of the data offset calculation at the beginning of a stage. The bus alignment is done in parallel with other tasks within a stage and therefore has a low overall performance impact. The pipeline alignment block microarchitecture is depicted in Figure~\ref{fig:pipe_align}.

This block delays the input data and performs bit-shifts to remove the already extracted data at the same parser stage. Shift amounts are functions of the header size and the bus size. In the case of fixed-size headers, these shift amounts are hardwired. For variable-sized headers, they are calculated by the \textit{ShiftAmount}, which is explained in more details in Section~\ref{sec:var_size}.

The output bus is then composed of data belonging to the current input data stream and from the previous cycle. When the current header instance is not to be processed, in the case where \textit{HeaderValidIn} is not set, this block just passes the input data to the output bus, playing the role of a bypass unit.

\subsubsection{Handling Variable-sized Headers}\label{sec:var_size}

It is not unusual to have a network protocol in which the header size is unknown until the packet arrives at a network equipment. The header size is inferred from a header field. IPv4 is such an example.


One approach to handle variable-sized headers would be to directly generate the required arithmetic circuit from the high-level packet processing program. However, this is an inefficient option based on our bus-aligned pipeline layout. In our architecture, supporting variable-sized would require dynamic barrel-shifters. Recall that a brute-force approach to design barrel-shifters uses a chain of multiplexers. For a N-bit barrel-shifter, this approach requires $N~log(N)$ multiplexers and introduces $log(N)$ combinational delay units to the critical path, compromising both FPGA resources and performance.

To get rid of dynamic barrel-shifters, we are inspired by a technique available in modern high-level programming languages known as template metaprogramming. Template metaprogramming uses the compiler capabilities to compute expressions at compilation time, improving the application performance. Based on this technique, during the P4 compilation in our framework, we calculate all valid results of arithmetic expressions storing them into ROM memories. These expressions include header size calculation and shift amount taps for static barrel-shifters. The results for a variable-sized IPv4 header instance showed 13\%~LUT and 15\%~FF usage reduction when implementing these ROM memories rather than dynamic barrel-shifters.


\subsection{Pipeline Layout Generation}\label{sec:pipe_gen}

\tikzset{%
  block/.style    = {draw, very thick, rectangle, minimum height = 2em,
    minimum width = 3em},
  sum/.style      = {draw, circle, node distance = 2cm}, 
  input/.style    = {coordinate}, 
  output/.style   = {coordinate} 
}

\tikzstyle{block} = [draw, rectangle, 
    minimum height=1em, minimum width=15em]
\tikzstyle{sum} = [draw, circle, node distance=1cm]
\tikzstyle{input} = [coordinate]
\tikzstyle{output} = [coordinate]
\tikzstyle{pinstyle} = [pin edge={to-,thin,black}]

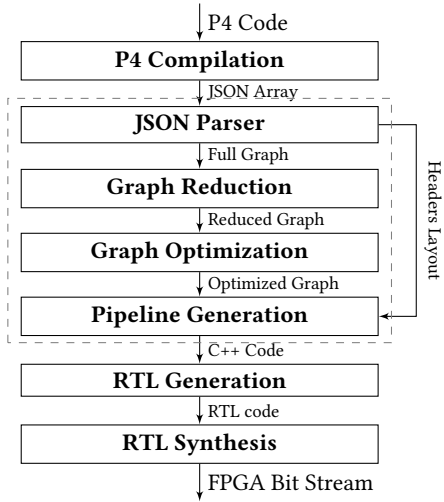
\begin{figure}[t]
\center
\begin{tikzpicture}[auto, node distance=1cm,>=latex']
    \node [input, name=input] {};
    \node [block, below of=input] (p4c) {\textbf{P4 Compilation}};
    \node [block, below of=p4c, node distance=.85cm] (json) {\textbf{JSON Parser}};
    \node [block, below of=json, node distance=.85cm] (graph_red) {\textbf{Graph Reduction}};
    \node [block, below of=graph_red, node distance=.85cm] (graph_opt) {\textbf{Graph Optimization}};
    \node [block, below of=graph_opt, node distance=.85cm] (pipe_gen) {\textbf{Pipeline Generation}};
    \node [block, below of=pipe_gen, node distance=.85cm] (rtl_gen) {\textbf{RTL Generation}};
    \node [block, below of=rtl_gen, node distance=.85cm] (rtl_synth) {\textbf{RTL Synthesis}};  
    \node [output, below of=rtl_synth] (output) {};

    \draw [draw,->] ($(p4c.north)+(0.0cm,.5cm)$) -- node {P4 Code} (p4c);
    \draw [->] (p4c) -- node {\footnotesize JSON Array} (json);
    \draw [->] (json.east) --+(0.5cm,0) |- node[pos=0.25, sloped, above] {\footnotesize Headers Layout} (pipe_gen);
    \draw [->] (json) -- node {\footnotesize Full Graph} (graph_red);
    \draw [->] (graph_red) -- node {\footnotesize Reduced Graph} (graph_opt);
    \draw [->] (graph_opt) -- node {\footnotesize Optimized Graph} (pipe_gen);
    \draw [->] (pipe_gen) -- node {\footnotesize C++ Code}(rtl_gen);
    \draw [->] (rtl_gen) -- node {\footnotesize RTL code}(rtl_synth);
    \draw [->] (rtl_synth) -- node {FPGA Bit Stream}($(rtl_synth.south)+(0.0cm,-.5cm)$);        
    \draw [color=gray,thin, dashed](-2.55, -4.75) rectangle (2.55,-1.5);
\end{tikzpicture}
\caption{Parser pipeline generation.}
\label{fig:gen_parser}
\end{figure}


\begin{figure*}[t]
\subfloat[][Original parser graph]
{
\begin{tikzpicture}[->,>=stealth',shorten >=1pt,auto,node distance=0.5cm, semithick, every fit/.style={rectangle, draw, inner sep=2pt}] \tikzstyle{every state}=[minimum size=1.3em, inner sep=1pt]

  \node[state]	    (Eth)  {\small ETH};
  \node[state]		(IPv4) [below left=of Eth] {\small IPv4};
  \node[state]		(IPv6) [below right=of Eth] {\small IPv6};
  \node[state]      (IPv6Ext) [below =of IPv6] {\small EXT};
  \node[state]		(UDP) [below =of IPv6Ext] {\small UDP};
  \node[state]		(TCP) at ($(IPv4.south |- UDP.west)$) {\small TCP};
  \node[state, accepting] (End) [below right =of TCP] {\small END};
    
  \path (Eth)  edge (IPv4)
          (Eth)  edge (IPv6)
          (Eth)  edge (End)
		  (IPv4) edge (UDP)          
		  (IPv4) edge (TCP)
		  (IPv4) edge [bend right=65] (End)
		  (IPv6) edge (IPv6Ext) 		  
		  (IPv6) edge[bend left=30] (UDP)          
		  (IPv6) edge (TCP)
		  (IPv6) edge[bend left=65] (End)
		  (IPv6Ext) edge (UDP)          
		  (IPv6Ext) edge (TCP)
		  (IPv6Ext) edge (End)
		  (UDP) edge (End)
		  (TCP) edge (End);

\end{tikzpicture}
\label{fig:graph:original_graph}
}\hfill
\subfloat[][Transitive reduction of the original graph]
{
\begin{tikzpicture}[->,>=stealth',shorten >=1pt,auto,node distance=0.5cm, semithick, every fit/.style={rectangle, draw, inner sep=2pt}] \tikzstyle{every state}=[minimum size=1.3em, inner sep=1pt]

  \node[state]	    (Eth)  {\small ETH};
  \node[state]		(IPv4) [below left=of Eth] {\small IPv4};
  \node[state]		(IPv6) [below right=of Eth] {\small IPv6};
  \node[state]      (IPv6Ext) [below =of IPv6] {\small EXT};
  \node[state]		(UDP) [below =of IPv6Ext] {\small UDP};
  \node[state]		(TCP) at ($(IPv4.south |- UDP.west)$) {\small TCP};
  \node[state, accepting] (End) [below right =of TCP] {\small END};
  
    \path (Eth)  edge (IPv4)
          (Eth)  edge (IPv6)
		  (IPv4) edge (UDP)          
		  (IPv4) edge (TCP)
		  (IPv6) edge (IPv6Ext) 		       
		  (IPv6Ext) edge (UDP)          
		  (IPv6Ext) edge (TCP)
		  (UDP) edge (End)
		  (TCP) edge (End);

\end{tikzpicture}
\label{fig:graph:reduced_graph}
}\hfill
\subfloat[][Equivalent reduced graph with an spare node]
{
\begin{tikzpicture}[->,>=stealth',shorten >=1pt,auto,node distance=0.5cm, semithick, every fit/.style={rectangle, draw, inner sep=2pt}] \tikzstyle{every state}=[minimum size=1.3em, inner sep=1pt]

  \node[state]	    (Eth)  {\small ETH};
  \node[state]		(IPv4) [below left=of Eth] {\small IPv4};
  \node[state, dotted]		(Dummy) [below =of IPv4] {\scriptsize dummy};
  \node[state]		(IPv6) [below right=of Eth] {\small IPv6};
  \node[state]      (IPv6Ext) [below =of IPv6] {\small EXT};
  \node[state]		(UDP) [below =of IPv6Ext] {\small UDP};
  \node[state]		(TCP) [below=of Dummy] {\small TCP};
  \node[state, accepting] (End) [below right=of TCP] {\small END};
  
    \path (Eth)  edge (IPv4)
          (Eth)  edge (IPv6)
		  (IPv4) edge (Dummy)  
		  (Dummy) edge (UDP)
		  (Dummy) edge (TCP)
		  (IPv6) edge (IPv6Ext) 		       
		  (IPv6Ext) edge (UDP)          
		  (IPv6Ext) edge (TCP)
		  (UDP) edge (End)
		  (TCP) edge (End);

\end{tikzpicture}
\label{fig:graph:dummy_graph}
}\hfill
\subfloat[][Final transformed graph]
{
\begin{tikzpicture}[->,>=stealth',shorten >=1pt,auto,node distance=0.5cm, semithick, every fit/.style={rectangle, draw, inner sep=2pt}] \tikzstyle{every state}=[minimum size=1.3em, inner sep=1pt]

  \node[state]	    (Eth)  {\small ETH};
  \node[state]		(IPv4) [below left=of Eth] {\small IPv4};
  \node[state]		(IPv6) [below right=of Eth] {\small IPv6};
  \node[state, dashed]      (IPv6Ext) [below left=of IPv6] {\small EXT};
  \node[state]		(UDP) [below right =of IPv6Ext] {\small UDP};
  \node[state]		(TCP) [below left=of IPv6Ext] {\small TCP};
  \node[state, accepting] (End) [below right=of TCP] {\small END};
  
    \path (Eth)  edge (IPv4)
          (Eth)  edge (IPv6)
		  (IPv4) edge (IPv6Ext)  
		  (IPv6) edge (IPv6Ext) 		       
		  (IPv6Ext) edge (UDP)          
		  (IPv6Ext) edge (TCP)
		  (UDP) edge (End)
		  (TCP) edge (End);

\end{tikzpicture}
\label{fig:graph:final_graph}
}\hfill
\caption{Parser graph transformation}
\label{fig:graph}
\end{figure*}
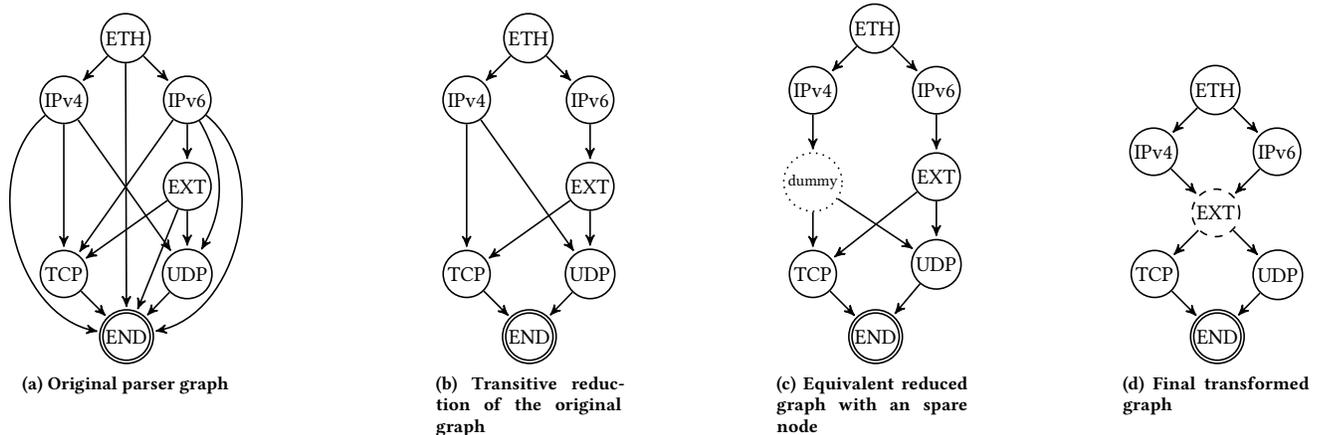

The procedure to generate the parser pipeline is depicted in Figure~\ref{fig:gen_parser}. The input P4 code is compiled using the P4C compiler \cite{p4c_git:16} producing a JSON array. We have chosen to use the result of the P4 back-end compilation (p4c-bm2-ss driver) for sake of simplicity.

Our work is limited to what is enclosed by the dashed rectangle in Figure~\ref{fig:gen_parser} and it is written in Python. It starts with the parsing of the JSON array file. While parsing, the script extracts the data structures necessary to initialize the multiple C++ \textit{Header} instances that compose the parser pipeline. The JSON parser also extracts the full parser graph. Figure~\ref{fig:graph:original_graph} presents a full parser graph generated from a header stack comprising the following protocols: Ethernet, IPv4, IPv6, IPv6 extension header, UDP and, TCP.

For an efficient pipelined design, the graph illustrated in Figure~\ref{fig:graph:original_graph} is not suitable. In that representation, almost all pipeline stages need bypass schemes to skip undesired state transitions, introducing combinational delays and increasing the resource usage due to the bypass multiplexers. We propose to simplify the original graph in order to have a more regular pipeline layout.

The graph simplification starts with the graph reduction phase that receives as input the full graph. This step performs a transitive reduction of the original graph in order to eliminate redundant graph edges. This phase also extracts the longest possible path of the parser graph. The result of this phase is shown in Figure~\ref{fig:graph:reduced_graph}.

The graph presented in Figure~\ref{fig:graph:dummy_graph} is an alternative representation for the reduced graph from Figure~\ref{fig:graph:reduced_graph}. In this graph, a dummy node is introduced to offer the same reachability while balancing the graph. This dummy node only acts as a bypass element and therefore has no implementation cost, thus, they can be merged with existent nodes at the same graph level.

\begin{algorithm}[t]
\caption{Graph balancing algorithm}
\label{alg:graph_opt}
\SetAlgoLined
\footnotesize
\SetKwFunction{graphBalance}{graphBalance}
\SetKwFunction{computeNodesLevel}{computeNodesLevel}
\SetKwFunction{successors}{successors}%
\SetKwFunction{removeEdge}{removeEdge}%
\SetKwFunction{addEdge}{addEdge}%
\SetKw{In}{in}%
\SetKw{Not}{not}%
\SetKw{Continue}{continue}%
\SetKwInOut{Input}{input}\SetKwInOut{Output}{output}
\Input{List of nodes representing a transitive reduced graph}
\Input{Ordered list of nodes belonging to the longest path}
\Output{Optimized balanced graph}
\KwData{A node is a data structure that has pointers to $successors$/$predecessors$ and methods to add/remove them. A node $level$ represents the graph level and it is unassigned at the beginning.}
\Fn{\graphBalance{tReducedGraph, longestPath}}{
	\tcc{Compute the distance of all nodes to the root}
	$\computeNodesLevel(tReducedGraph)$\label{alg:comp_level}

	\tcc{Remove edges to successors from nodes not in the longest path}
	\For{$node$ \In $tReducedGraph$}{\label{alg:del_succ_ini}
		\If{$node \not\in longestPath$}{
			\For{$sucNode$ \In $node.\successors()$}{
				$\removeEdge(node, sucNode)$
			}
		}
	}\label{alg:del_succ_end}

	\tcc{Adding spare edges to balance the graph}
	\For{$node$ \In $tReducedGraph$}{\label{alg:add_succ_ini}
		\If{$node \not\in longestPath$}{
			$\addEdge(node, longestPath[node.level+1])$
		}
	}\label{alg:add_succ_end}
	\Return $tReducedGraph$\label{alg:return_graph}
}
\end{algorithm}

We propose a graph balancing algorithm in Algorithm~\ref{alg:graph_opt} to optimize the reduced graph. It receives as parameters the transitive reduced parser graph and the longest path in the graph. As output, the algorithm returns a balanced graph tailored to our pipelined architecture. The first function call (line~\ref{alg:comp_level}) in the algorithm executes the node level computation in relation to the root for all nodes. The first loop (lines~\ref{alg:del_succ_ini}~-~\ref{alg:del_succ_end}) iterates over the nodes that are not in the longest path. It deletes the edges from these nodes to their children. The last loop (lines~\ref{alg:add_succ_ini}~-~\ref{alg:add_succ_end}) iterates again over the nodes that are not part of the longest path and assigns a child to them. The chosen child is the first one belonging to the next graph level. Finally, the algorithm returns an optimized graph on line~\ref{alg:return_graph}. An example of balanced graph is shown in Figure~\ref{fig:graph:final_graph}.

The last step of the proposed approach illustrated in Figure~\ref{fig:gen_parser} is the code generation. This phase receives as input a set of data structures representing the supported header layouts and the balanced graph. The header layouts are used to initialize both template and construction parameters for the C++ objects. The pipeline layout is drawn based on the balanced graph, with multiplexer insertion when required. 
The result of this phase is a synthesizable C++ code.

The generated C++ code is tailored for FPGA implementation. The next step in the processing chain is to generate RTL code for FPGA synthesis and place-and-route. Vivado HLS 2015.4 is used in this phase. Then, the generated RTL is synthesized under Vivado, which produces a bit stream file compatible with Xilinx FPGAs.

\section{Experimental Results}\label{sec:results}

To demonstrate and evaluate our proposed method, we conducted two classes of experiments, the same ones performed in \cite{Benacek:2016}, to simplify comparisons. These two classes are defined as follows:
\begin{itemize}[noitemsep,topsep=0pt]
\item \textbf{Simple parser}: Ethernet, IPv4/IPv6 (with 2 extensions), UDP, TCP, and ICMP/ICMPv6; and
\item \textbf{Full parser}: same as simple parser plus MPLS (with two nested headers) and VLAN (inner and outer).
\end{itemize}

We used Vivado HLS 2015.4 to generate synthesizable RTL code. The RTL code was afterwards synthesized under Vivado 2015.4. The target FPGA device of this work was a Xilinx Virtex-7 FPGA.

\begin{table*}[t]
\centering
\caption{Parser results comparison}
\label{tab:parser_resutls}
\begin{threeparttable}
\begin{tabular}{|c|S[table-format=3.0]|S[table-format=3.1]|S[table-format=3.0]|S[table-format=2.1]|S[table-format=5.0]|S[table-format=5.0]|S[table-format=5.0]|c|}
\hline
\multicolumn{1}{|c|}{\multirow{3}{*}{\textbf{Work}}} & \multicolumn{4}{c|}{\textbf{Performance}}                                                            & \multicolumn{3}{c|}{\textbf{Resources}} &\multicolumn{1}{c|}{\multirow{3}{*}{\textbf{\makecell{Extracted \\ Fields}}}}                  \\ \cline{2-8}
\multicolumn{1}{|c|}{}                               & \textbf{Data Bus} & \textbf{Frequency} & \textbf{Throughput} & \textbf{Latency} & \textbf{LUTs} & \textbf{FFs} & \textbf{Slice Logic} & \\

\textbf                             & \textbf{{[}bits{]}} & \textbf{{[}MHz{]}}         & \textbf{{[}Gb/s{]}}  & \textbf{{[}ns{]}} &                  &                & \textbf{(LUTs+FFs)} & \\ 
\hline
\multicolumn{9}{|c|}{\textbf{Simple Parser}}                                                                                                                                                \\ \hline
\cite{Gibb:2013}              & 256                 & 184.1                    & 47               & N/A               & 14906            & 2963           & 17869  & All fields             \\
\cite{Gibb:2013}             & 256                 & 178.6                    & 46               & N/A               & 6865             & 1851           & 8716   & TCP/IP 5-tuple                          \\
Golden~\cite{Benacek:2016}         & 512                 & 195.3         & 100                  & 15       & N/A              & N/A       & 5000 & TCP/IP 5-tuple    \\ 
\cite{Benacek:2016}          & 512                 & 195.3         & 100                  & 29       & N/A              & N/A       & 12000 & TCP/IP 5-tuple    \\ 
Hybrid \cite{Benacek:2016} and this work                   & 320                 & 312.5                      & 100                  & 28.8              & 4699             & 7254       & 11953  & TCP/IP 5-tuple \\ 
This work                    & 320                 & 312.5                      & 100                  & 19.2              & 4270             & 6163       & 10433  & TCP/IP 5-tuple
\\
This work                     & 320                 & 312.5                      & 100                  & 19.2              & 5888             & 10448      & 16336  & All fields             \\ 
\hline
\multicolumn{9}{|c|}{\textbf{Full Parser}}                                                                                                                                                  \\ \hline
\cite{Gibb:2013}              & 64                  & 172.2                    & 11               & N/A               & 6946             & 2600           & 9546     & All fields            \\
\cite{Gibb:2013}             & 64                  & 172.2                    & 11               & N/A               & 3789             & 1425           & 5214     & TCP/IP 5-tuple           \\ 
Golden~\cite{Benacek:2016}          & 512                 & 195.3    & 100                  & 27              & N/A            & N/A           & 8000 & TCP/IP 5-tuple              \\ 
\cite{Benacek:2016}          & 512                 & 195.3         & 100                  & 46.1              & 10103            & 5537           & 15640 & TCP/IP 5-tuple              \\ 
Hybrid \cite{Benacek:2016} and this work                    & 320                 & 312.5                      & 100                  & 41.6                & 6450             & 10308          & 16758  & TCP/IP 5-tuple             \\
This work                   & 320                 & 312.5                      & 100                  & 25.6                & 6046             & 8900          & 14946  & TCP/IP 5-tuple             \\ 
This work                    & 320                 & 312.5                      & 100                  & 25.6                & 7831             & 13671          & 21502  & All fields              \\
\hline
\end{tabular}
\end{threeparttable}
\end{table*}

Table~\ref{tab:parser_resutls} shows a comparison against others works present in the literature \cite{Gibb:2013,Benacek:2016} that support fixed- and variable-sized headers. In the case of \cite{Gibb:2013}, because they do not provide FPGA results, we reproduced their results based on a framework provided by the authors \cite{parser_git:16}. For that, we developed a script that converts the P4 code to the data structures needed in the framework. 

Analysing the data from Table~\ref{tab:parser_resutls}, both this work and \cite{Benacek:2016} outperform \cite{Gibb:2013}, which is expected since the framework proposed in that work for automatic parser generation was designed for ASIC implementation and not for FPGA.

We assume as a golden model, labelled as Golden \cite{Benacek:2016} in Table~\ref{tab:parser_resutls}, a hand-written VHDL implementation presented in \cite{Benacek:2016}, which the authors used to evaluate their method.

Under the same design constraints, our work achieves the same throughput as \cite{Benacek:2016}, while not only reducing latency by 45\% but also the LUT consumption by 40\%. However, our architecture consumes more FFs, which is partially explained by the additional pipeline registers inferred by the Vivado HLS. Nonetheless, we can even have a lower overall slice utilization compared to \cite{Benacek:2016}, since in a Virtex-7 each slice has four LUTs and eight FFs, and our architecture does not double the number of used FFs.

Also, a notable resource consumption reduction is noticed when the number of extracted fields are reduced from all fields to 5-tuple, since a large amount of resources is destined to store the extracted fields, which matches with the findings reported in \cite{Gibb:2013}.

To compare the impact of our proposed pipelined layout, we implemented the pipeline organization proposed in \cite{Benacek:2016} using the proposed header block architecture illustrated in Figure~\ref{fig:high_level:header} since their source code was unavailable. This experiment is marked as "Hybrid \cite{Benacek:2016} and this work" in Table~\ref{tab:parser_resutls}. For the simple parser, our proposed architecture improves latency by more than 33\%, while reducing by 16\% and 10\% number of used FFs and LUTs, respectively. In the case of the full parser, the latency was reduced by 39\%, while the resource consumption follows the results of the simple parser.

Moreover, this hybrid solution also outperforms the original work \cite{Benacek:2016} in both latency and LUT consumption. It shows that our microarchitectural choices are more efficient in these aspects. In addiction, these better results can also be related to the language chosen to describe each architecture. In \cite{Benacek:2016}, they generated VHDL code from a P4 description. Our design uses templated C++ classes, which can fill the abstraction gap between the high-level packet processing program and the low-level RTL code.


When comparing to the golden model, the results obtained with our architecture are comparable to it in terms of latency. Our design, however, utilizes nearly twice the overall amount of logic resources, following what has been reported in \cite{Benacek:2016}. However, since separate LUTs and FFs, or slice consumption results are unavailable, we cannot fairly compare resources utilization results.

\begin{figure}[ht]
\centering
\begin{tikzpicture}
    \begin{axis}[
	   	legend columns=2,
    		footnotesize,
    		width = .5\textwidth,
    		height=0.275\textwidth,
        ybar=5*\pgflinewidth,
        bar width=0.015\textwidth,
        ymajorgrids = true,
        ylabel = {LUTs/FFs (Thousands)},
        xlabel = {Throughput [Gb/s]},
        symbolic x coords={10,40,100,160},
        xtick = data,
        scaled y ticks = false,
        enlarge x limits=0.25,
        axis line style={-},
        ymin=0,ymax=20,
        area legend,
        legend pos=north west,
    ]
        \addplot[style={mark=none}]
           coordinates {(10, 4.156) (40, 5.230) (100, 5.888) (160, 7.372) };
        \addplot[style={pattern=dots}]
           coordinates {(10, 4.927) (40, 6.439) (100, 7.831) (160, 10.181) };
        \addplot[style={pattern=north east lines}]
           coordinates {(10, 6.551) (40, 9.411) (100, 10.448) (160, 13.822) };
        \addplot[style={pattern=grid}]
           coordinates {(10, 7.959) (40, 12.177) (100, 13.671) (160, 19.767) };
        \legend{\scriptsize LUTs (simple parser), \scriptsize LUTs (full parser), \scriptsize FFs (simple parser), \scriptsize FFs (full parser)}
    \end{axis}
\end{tikzpicture}
\caption{Synthesis results for multiple data rate parsers.}
\label{fig:arch_scale}
\end{figure}
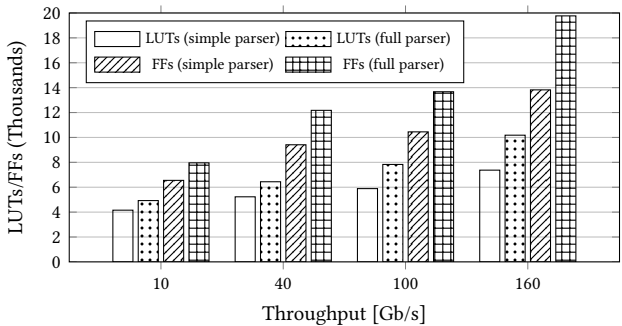

As shown in Table~\ref{tab:parser_resutls}, the present work achieves the best maximum frequency comparing to state-of-the-art, which allows scaling to data rates higher than \SI{100}{\giga\bit/\second}. Figure~\ref{fig:arch_scale} presents the design scalability results for data rates ranging from \SI{10}{\giga\bit/\second} up to \SI{160}{\giga\bit/\second}. It is worth noting that the data rate scaling causes a non-expressive impact in terms of LUTs, corresponding to an increase of $35~\sfrac{\text{LUTs}}{\text{Gbps}}$ in the case of the full \SI{160}{\giga\bit/\second} parser.


\section{Conclusion}\label{sec:conclusion}


FPGAs have increasingly gained importance in today's network equipment. FPGAs provide flexibility and programmability required in SDN-based networks. SDN-aware FEs need to be reconfigured to be able to parse new protocols that are constantly being deployed.

In this work, we proposed an FPGA-based architecture for high-speed packet parsing described in P4. Our architecture is completely described in C++ to raise the development abstraction. Our methodology includes a framework for code generation, including a graph reducing algorithm for pipeline simplification. From modern high-level languages, we borrowed the idea of metaprogramming to perform offline expressions calculation, reducing the burden of calculating them at run-time.

Our architecture performs as well as the state-of-the-art while reducing latency and LUT usage. The latency is reduced by 45\% and the LUT consumption is reduced by 40\%. Our proposed methodology allows a throughput scalability ranging from \SI{10}{\giga\bit/\second} up to \SI{160}{\giga\bit/\second}, with moderate increasing in logic resources usage.


\begin{acks}
The authors thank A. Abdelsalam, M. D. Souza Dutra, I. Benacer, T. Stimpfling, and T. Luinaud for their comments. This work is supported by the CNPQ-Brazil.
\end{acks}

\balance

\bibliographystyle{ACM-Reference-Format}

\bibliography{sample-bibliography} 

\end{document}